\documentclass[9pt,twocolumn,twoside]{opticajnl}
\journal{opticajournal}
\setboolean{shortarticle}{true}
\setboolean{displaycopyright}{false}
\usepackage{amsmath,amssymb}
\usepackage{microtype}
\usepackage{placeins}

\renewcommand{\eqref}[1]{\textup{(\ref{#1})}}
\hypersetup{pdftitle={A fast and accurate model for gradient-based inverse design of multimode waveguides},pdfauthor={}}
\title{A fast and accurate model for gradient-based inverse design of multimode waveguides}
\author[1]{Yuanze Chen}
\author[1]{Zheng-Hua Chen}
\author[1,2*]{Yan-Xiao Gong}
\author[1]{Shi-Ning Zhu}
\affil[1]{National Laboratory of Solid State Microstructure, School of Physics, and Collaborative Innovation Center of Advanced Microstructures, Nanjing University, Nanjing 210093, China}
\affil[2]{Jiangsu Key Laboratory of Quantum Information Science and Technology, Nanjing University, Nanjing 210093, China}
\affil[*]{gongyanxiao@nju.edu.cn}
\date{}
\dates{}
\doi{}

\begin{abstract}
Multimode photonic inverse design requires fast, accurate computational models that provide geometry gradients. Based on coupled-mode theory (CMT), we propose a differentiable surrogate model. Its parameter-extraction method uses a change of modal basis to obtain coupled-mode parameters from cross-sectional mode data. We also provide a complete, easy-to-implement propagation and optimization workflow. Using finite-difference time-domain (FDTD) simulation data as a common reference, we compare the prediction accuracy of our model and forward dataset-based eigenmode expansion (EME) for randomly generated thin-film lithium niobate (TFLN) waveguide profiles that traverse modal coupling regions. Using identical modal libraries, the model reduces the maximum per-mode mean absolute transmission error by $17$--$37\%$ compared with forward EME. We demonstrate a spot-size converter design that preserves the fundamental TM$_0$ mode in its hybridization region with the first-order TE$_1$ mode and obtain nine structures, each with FDTD-verified transmission above $98\%$. These tests demonstrate the model's utility for large-scale photonic inverse design, enabling rapid exploration of the design parameter space by combining gradient information with improved prediction accuracy.

\end{abstract}

\begin{document}
\maketitle
\raggedbottom

\begin{figure*}[!t]
  \centering
  \includegraphics[width=\textwidth]{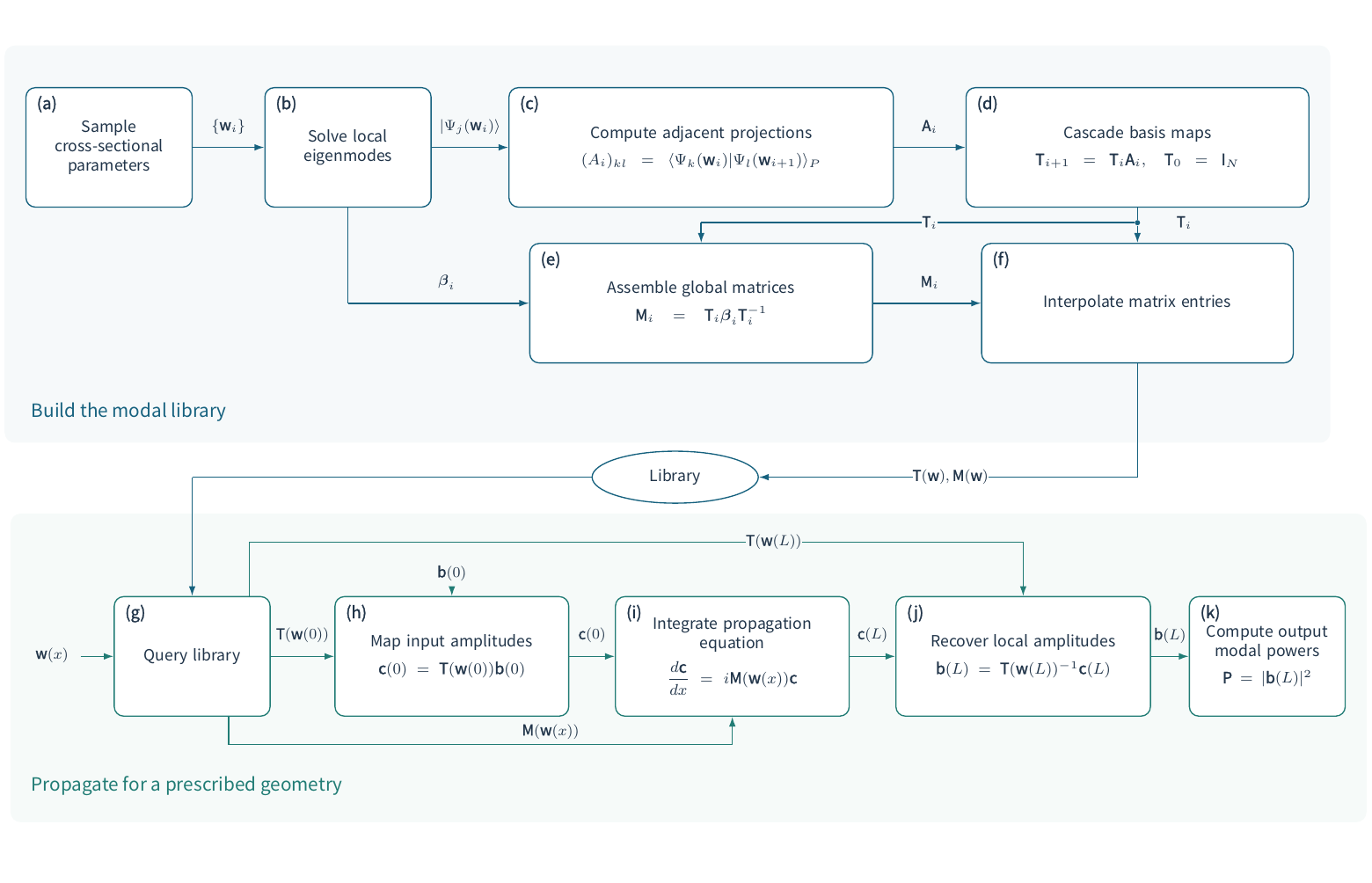}
  \caption{Global coupled-mode workflow. (a--f) Modal-library construction. (g--k) Propagation for a prescribed geometry.}
  \label{fig:workflow}
\end{figure*}

\begin{figure*}[!t]
  \centering
  \includegraphics[width=0.95\textwidth]{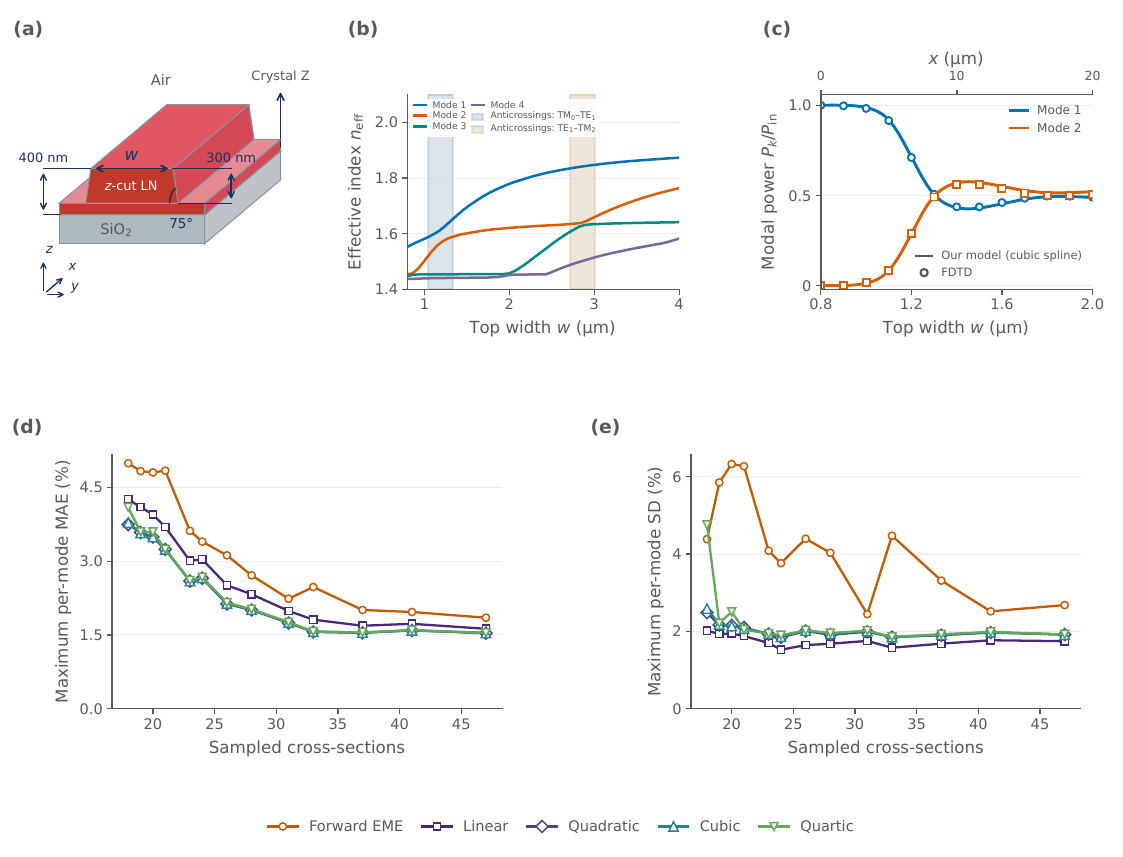}
  \caption{Waveguide geometry and propagation benchmarks. (a) TFLN ridge-waveguide geometry. (b) Effective indices of the four retained local modes. (c) Local-mode powers along a linear taper under TM$_0$ excitation. (d,e) Maximum per-mode MAE and SD, respectively, of absolute transmission errors relative to FDTD over modes 1--3 ($n=100$ random waveguides). Wavelength: $1.55\,\mu$m. Powers and errors are normalized to incident power.}
  \label{fig:comparison}
\end{figure*}

\begin{figure*}[!t]
  \centering
  \includegraphics[width=0.83\textwidth]{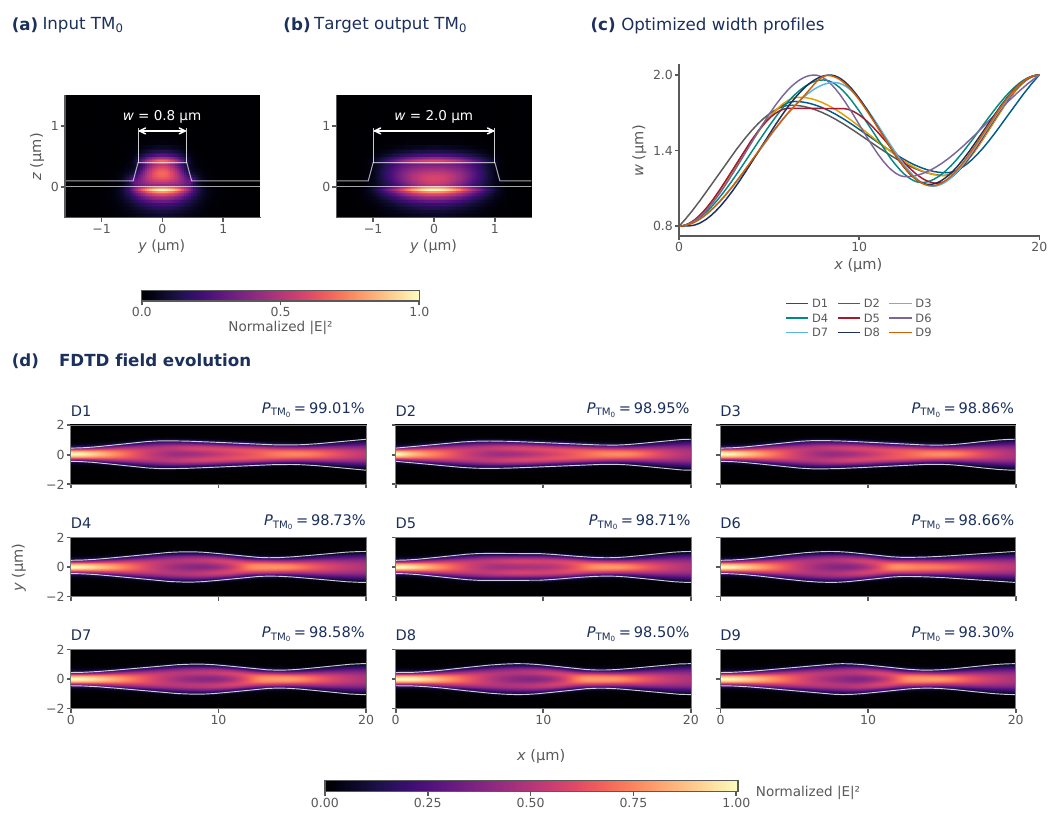}
  \caption{Optimized TM$_0$-preserving spot-size converters. (a,b) Input and target output eigenmode intensities. (c) Width profiles of D1--D9. (d) FDTD intensities in an $x$--$y$ plane inside the waveguide under TM$_0$ excitation; white curves mark the waveguide boundaries. $P_{\mathrm{TM}_0}$ denotes output TM$_0$ power normalized to incident power.}
  \label{fig:optimization}
\end{figure*}

Inverse design enables compact photonic devices with improved performance, but requires a large number of electromagnetic simulations~\cite{molesky2018inverse,piggott2015inverse,shen2015integratednanophotonics}. Gradient-based optimization can substantially accelerate this search by using objective-function gradients to guide geometry updates~\cite{lalau-keraly2013adjoint}. Supporting gradient-based optimization therefore requires computational models that provide geometry gradients as well as fast predictions. Conventional finite-difference time-domain (FDTD) simulations are computationally expensive, and their cumulative cost can become prohibitive in large-scale design tasks that require many simulations~\cite{christiansen2021inverse,hammond2022highperformance}. To rapidly explore the design parameter space before high-accuracy FDTD validation~\cite{campbell2019review,yeung2023deepadjoint}, dataset-based eigenmode expansion (EME) has been proposed as a fast surrogate model that reuses precomputed cross-sectional modes and avoids repeated, computationally intensive mode-overlap calculations~\cite{oktay2024computationally,song2025ultrafast}. It evaluates propagation by matching fields between successive cross-sections~\cite{gallagher2003eigenmode,bienstmanoptical}. However, selecting discrete entries from the library makes it difficult to compute gradients with respect to geometry.

Our model comprises a theoretical formulation of intermodal energy exchange, a parameter-extraction method, and a differentiable propagation and optimization workflow. In the theoretical formulation, we express the coupled-mode equations in a fixed reference basis. By transforming the local modes to a common global basis, we map modal data computed at discrete cross-sections into the parameters of a single set of coupled-mode equations. Interpolating these parameters can reduce errors caused by discrete cross-sectional sampling. It also represents the coupled-mode parameters as continuous, piecewise differentiable functions of geometry, enabling direct computation of geometry gradients. The resulting model combines accurate and fast propagation predictions with direct support for gradient-based optimization.

We consider lossless, reciprocal waveguides and neglect back-reflection. We set the waveguide propagation direction along $x$, the width direction along $y$, and the vertical direction along $z$. As in numerical EME, we use a truncated expansion in local eigenmodes~\cite{oktay2024computationally}. At each longitudinal position $x$, we retain the $N$ forward guided modes with the highest effective indices. We assume that coupling to omitted modes is negligible. We write each local eigenmode as $|\Psi_j(x)\rangle=[\hat{\mathbf e}_{tj},\hat{\mathbf h}_{tj}]^T$, where $\hat{\mathbf e}_{tj}$ and $\hat{\mathbf h}_{tj}$ are its unit-power electric and magnetic field profiles. The subscript $t$ denotes components transverse to $x$, $j=1,\ldots,N$ labels modes in descending order of effective index, and the superscript $T$ denotes transpose. We define modal projection through the power inner product:

\begin{equation}
 \langle\Psi_k|\Psi_l\rangle_P\equiv\frac14\int
 (\hat{\mathbf e}_{tk}^{*}\!\times\hat{\mathbf h}_{tl}
 +\hat{\mathbf e}_{tl}\!\times\hat{\mathbf h}_{tk}^{*})\cdot\hat{\mathbf x}\,dA
 =\delta_{kl}.
 \label{eq:power}
\end{equation}
Here $k,l=1,\ldots,N$ are mode indices, the subscript $P$ denotes the power inner product, $*$ denotes complex conjugation, and $\delta_{kl}$ is the Kronecker delta. The integral is over the transverse $y$--$z$ plane, $dA=dy\,dz$ is its area element, and $\hat{\mathbf x}$ is the unit vector along propagation.

In this local basis, the amplitude vector $\mathbf b$ obeys the coupled-mode equations $d\mathbf b/dx=(i\boldsymbol\beta+\mathbf C)\mathbf b$, where $\boldsymbol\beta=\operatorname{diag}(\beta_j)$ contains the propagation constants $\beta_j$ obtained by solving the local eigenmode problem, and the coupling matrix $\mathbf C$ has entries $C_{kl}=-\langle\Psi_k|\partial_x\Psi_l\rangle_P$~\cite{snyder1983optical,johnson2002adiabatic,huang1990selfconsistent,huang1994coupledmode,siriani2021adiabatic}. The basis transformation $T_{mj}=\langle\Phi_m|\Psi_j\rangle_P$ projects onto fixed reference modes $|\Phi_m\rangle$, with $m=1,\ldots,N$. We assume approximate completeness, $P_N(x)=\sum_{k=1}^{N}|\Psi_k\rangle\langle\Psi_k|_P\approx I$. Under this assumption, we obtain $\mathbf C=-\mathbf T^{-1}(d\mathbf T/dx)$. In terms of the reference-basis amplitudes $\mathbf c=\mathbf T\mathbf b$, the basis-derivative terms cancel, giving the following equation with global propagation matrix $\mathbf M$ and parameter vector $\mathbf w(x)=[w_1(x),\ldots,w_p(x)]^T$ describing the waveguide cross-sectional shape:

\begin{equation}
 \frac{d\mathbf c}{dx}=i\mathbf M(\mathbf w(x))\mathbf c,
 \qquad \mathbf M=\mathbf T\boldsymbol\beta\mathbf T^{-1}.
 \label{eq:global}
\end{equation}

Here $\mathbf w$ represents width and other cross-sectional parameters. Propagation in this basis requires only the entries of $\mathbf M$, which can be interpolated as continuous functions of $\mathbf w$ for typical waveguide design problems.

We first build a modal library by sampling the cross-sectional parameter domain and solving for the $N$ retained eigenmodes at each sample [Fig.~\ref{fig:workflow}(a,b)]. We order the modes by decreasing effective index and choose the modes at a sampled reference cross-section $\mathbf w_{\rm ref}$ as the fixed basis. Starting from this reference at $i=0$, we extract $\mathbf T$ by cascading projections between adjacent cross-sections [Fig.~\ref{fig:workflow}(c,d)]. Under completeness, this cascade is equivalent to direct projection onto the fixed reference basis:

\begin{equation}
 \begin{aligned}
  \mathbf T_{i+1}&=\mathbf T_i\mathbf A_i,\qquad \mathbf T_0=\mathbf I_N,\\
  (A_i)_{kl}&=\langle\Psi_k(\mathbf w_i)|\Psi_l(\mathbf w_{i+1})\rangle_P.
 \end{aligned}
 \label{eq:maps}
\end{equation}

Here $\mathbf w_i$ denotes the parameters of sample $i$, $\mathbf A_i$ projects the modes at sample $i+1$ onto those at sample $i$, and $\mathbf I_N$ is the $N\times N$ identity matrix. We form $\mathbf M$ at each sample and interpolate its entries over $\mathbf w$ [Fig.~\ref{fig:workflow}(e,f)].

For a prescribed geometry $\mathbf w(x)$, interpolation of the sampled matrices gives $\mathbf M(x)=\mathbf M(\mathbf w(x))$ at any longitudinal position $x$ [Fig.~\ref{fig:workflow}(g)]. For arbitrary local input amplitudes $\mathbf b(0)$, we set $\mathbf c(0)=\mathbf T(\mathbf w(0))\mathbf b(0)$ [Fig.~\ref{fig:workflow}(h)]. We integrate Eq.~\eqref{eq:global} and recover the local amplitudes as $\mathbf b(x)=\mathbf T(\mathbf w(x))^{-1}\mathbf c(x)$ [Fig.~\ref{fig:workflow}(i,j)]. The output energy distribution among the modes for this initial condition is represented by the modal-power vector $\mathbf P=|\mathbf b(L)|^2$ [Fig.~\ref{fig:workflow}(k)], where the absolute square is taken componentwise and $L$ is the propagation distance between the input and output cross-sections.

For geometry optimization, let $\mathcal F(\mathbf c(L))$ be a real objective and keep the ports and integration interval fixed. We solve the adjoint equation $d\boldsymbol\eta/dx=i\mathbf M^\dagger\boldsymbol\eta$ backward from $\boldsymbol\eta(L)=\partial\mathcal F/\partial\mathbf c(L)^*$. For a real design parameter $\theta_a$, the gradient is

\begin{equation}
 \frac{d\mathcal F}{d\theta_a}=2\operatorname{Re}\int_0^L\sum_{r=1}^{p}
 \boldsymbol\eta^\dagger i\frac{\partial\mathbf M}{\partial w_r}
 \mathbf c\,\frac{\partial w_r}{\partial\theta_a}\,dx.
 \label{eq:gradient}
\end{equation}

Here $w_r$ ($r=1,\ldots,p$) are the components of $\mathbf w$, and $\partial\mathbf M/\partial w_r$ is obtained directly from the interpolated operator without differentiating the eigenmode solver. In practice, these geometry gradients can be computed using automatic differentiation.

To demonstrate how the model works and benchmark its performance, we use the air-clad, $z$-cut thin-film lithium niobate (TFLN) ridge waveguide on silica shown in Fig.~\ref{fig:comparison}(a). We build the modal library at $\lambda=1.55\,\mu$m. The film is $400\,$nm thick, the etch depth is $300\,$nm, and the sidewalls make an angle of $75^\circ$ with the substrate plane. The crystal $Z$ axis is normal to the chip. The waveguide structure, with its sloped sidewalls, air cladding, and silica substrate, has pronounced vertical asymmetry, which contributes to strong mode hybridization~\cite{kaushalram2020mode}. Only the top width $w$ varies. The library samples widths of $0.8$--$4.0\,\mu$m at $10\,$nm intervals and retains $N=4$ modes. The modes of the $0.8\,\mu$m-wide cross-section define the reference basis.

For the TFLN waveguide structure described above, two anticrossings, involving TM$_0$--TE$_1$ and TE$_1$--TM$_2$, occur within the top-width range $w=0.8$--$4.0\,\mu$m [Fig.~\ref{fig:comparison}(b)], leading to strong power transfer among these modes~\cite{kaushalram2020mode,chen2025polarization}. To directly test the model's ability to predict intermodal power transfer, Fig.~\ref{fig:comparison}(c) shows the power evolution of the first two local modes under TM$_0$ excitation in a $20\,\mu$m-long taper whose top width increases linearly from $0.8$ to $2.0\,\mu$m. The pronounced power exchange confirms strong mode hybridization over this interval. Our model with cubic-spline interpolation accurately reproduces the FDTD benchmark, demonstrating its validity for a representative taper through a hybridization region.

Beyond a single monotonic taper, assessing surrogate accuracy for more varied geometries requires systematic statistical tests at different sampling densities. We vary the width step between adjacent sampled cross-sections to generate 13 distinct sets of modal-overlap data. Forward EME and our model use the same data from each set, with our model using four interpolation schemes for $\mathbf M$: linear interpolation and quadratic, cubic, and quartic splines. To benchmark these five surrogate models, we interpolate randomly generated control points to obtain 100 waveguide profiles that traverse the hybridization regions, with lengths of approximately $17$--$60\,\mu$m. We limit the rate of width variation to reduce leakage outside the retained guided-mode space and maintain the completeness assumption. Three-dimensional FDTD simulations provide reference output transmissions into the four retained modes. For each mode, we compute the mean absolute error (MAE) and population standard deviation (SD) of the absolute transmission errors over the 100 devices. Figures~\ref{fig:comparison}(d,e) report the respective maxima over the first three modes, quantifying prediction accuracy and device-to-device error variation. Both metrics are expressed as percentages of incident power.

For all surrogate models, the MAE curves decrease overall as sampling density increases and become nearly flat at 37 sampled cross-sections, indicating numerical convergence [Fig.~\ref{fig:comparison}(d,e)]. The higher-degree schemes give similar results, with lower MAE than linear interpolation and comparable SD, except for the increased quartic-spline SD at the lowest sampling density (18 cross-sections). This increase indicates that a higher interpolation degree can compromise stability when sampling is too sparse. Among the higher-degree interpolation schemes with lower MAE, we select the commonly used cubic spline for the following analysis.

Quantitatively, our cubic-spline model delivers higher accuracy and greater robustness than forward EME across all 13 modal libraries. It reduces the maximum per-mode MAE by $17$--$37\%$ relative to forward EME. Forward EME reaches $1.850\%$ with 47 cross-sections, whereas our model reaches $1.752\%$ with only 31, requiring approximately $34\%$ fewer sampled cross-sections. The cubic-spline model also yields lower SD across all libraries, corresponding to smaller device-to-device variation in absolute transmission errors. Its SD curve fluctuates less with sampling density, indicating reduced sensitivity to discrete cross-sectional sampling. The improved accuracy with fewer sampled cross-sections reduces the computational cost of constructing the modal library.

Building on the model's accuracy across diverse geometries, we use a device-design example to demonstrate how it supports large-scale gradient-based optimization. The example addresses polarization-preserving spot-size conversion in the hybridization region, a task hindered by the TM$_0$--TE$_1$ coupling discussed above~\cite{guo2015analysis}. In the adiabatic limit, input TM$_0$ follows the hybridized eigenmode into output TE$_1$, transferring its power to TE$_1$~\cite{dai2012mode,wang2021efficient}. Our design objective is therefore to expand the mode while retaining as much power as possible in TM$_0$, thereby preserving the input polarization.

For the spot-size converter, we choose input and output widths of $0.8$ and $2.0\,\mu$m, respectively, on either side of the TM$_0$--TE$_1$ hybridization region [Fig.~2(b)], and set the maximum device length to $20\,\mu$m. The longitudinal positions and widths of three internal nodes provide six design parameters. A seventh sets the taper length. The node widths lie between the port widths. These parameters define a multiparameter width profile that can be freely optimized, vary nonmonotonically, and remain within the library range. We minimize

\begin{equation}
 \mathcal J=-P_{\mathrm{TM}_0}
 +10\sum_i\left[\max\left(0,\left|\frac{w_{i+1}-w_i}{x_{i+1}-x_i}\right|-0.15\right)\right]^2,
 \label{eq:objective}
\end{equation}
where $P_{\mathrm{TM}_0}$ is the normalized transmitted power in the output TM$_0$ mode under TM$_0$ excitation. The penalty limits the rate of width change between neighboring nodes to suppress scattering and power leakage caused by rapid geometric variations. We randomly generate 4096 starting geometries, providing diverse initial structures for gradient-based optimization. We optimize each structure using the limited-memory Broyden--Fletcher--Goldfarb--Shanno algorithm~\cite{liu1989limited} until stagnation.

From the 4096 optimization runs, we ultimately obtain 30 shape groups. We select the profile with the highest predicted transmission in each group and evaluate with FDTD the 14 representatives whose surrogate-predicted transmission exceeds $95\%$. Nine achieve TM$_0$ transmission of $98.30$--$99.01\%$ [Fig.~\ref{fig:optimization}]. Figure~\ref{fig:optimization}(c) directly shows the width profiles of these nine structures. Their common broadening--narrowing--expanding trend may result from the limited number of nodes. Differences in their field evolution are shown in Fig.~\ref{fig:optimization}(d).

In the present optimization example, searches from 4096 starting geometries required on the order of $10^5$ surrogate evaluations and finished in approximately two hours on a desktop CPU. Building the offline library takes approximately $10$ hours; repeated evaluations share this cost. Using FDTD for this many evaluations would be prohibitively expensive. Our model thus offers a clear speed advantage over FDTD. This demonstrates the model's ability to handle large-scale gradient-based optimization tasks.

The current formulation describes forward propagation of lossless guided modes. With suitable modifications, the formulation could also accommodate back-reflection, modes computed with perfectly matched layers, and lossy modes. Although this example considers only width as a cross-sectional design variable, the model could support multiple geometric variables using the dataset-construction strategy of Ref.~\cite{song2025ultrafast}.

In conclusion, our model with cubic-spline interpolation improves transmission-prediction accuracy and robustness to library sampling density relative to forward EME, directly provides geometry gradients, and retains a computational cost far below FDTD. The converter design task further demonstrates its utility for large-scale gradient-based optimization.
\begin{backmatter}
\bmsection{Funding}
Natural Science Foundation of Jiangsu Province (BK20243060); National Natural Science Foundation of China (No. 62288101).
\smallskip

\bmsection{Disclosures} The authors declare no conflicts of interest.

\bmsection{Data Availability Statement} Data underlying the results can be obtained from the authors upon reasonable request.

\end{backmatter}
\FloatBarrier
\bibliography{references}
\clearpage

\end{document}